\documentclass{article}
\usepackage{graphicx}
\begin{document}

\title{Memory Based Collaborative Filtering with Lucene}

\author{Claudio Gennaro \\ claudio.gennaro@isti.cnr.it \\ ISTI-CNR, Pisa, Italy}
\maketitle
\begin{abstract}
Memory Based Collaborative Filtering is a widely used approach to provide recommendations. It exploits similarities between ratings across a population of users by forming a weighted vote to predict unobserved ratings. Bespoke solutions are frequently adopted to deal with the problem of high quality recommendations on large data sets. A disadvantage of this approach, however, is the loss of generality and flexibility of the general collaborative filtering systems. In this paper, we have developed a methodology that allows one to build a scalable and effective
collaborative filtering system on top of a conventional full-text search engine such as Apache Lucene.

\end{abstract}

\section{Introduction}
\label{sec:intro}

Collaborative Filtering (CF) is the task of predicting unknown rates of a given user for some items (e.g., products, movies) by taking into account other users' preferences.
One of the most commonly employed mechanism to approach the problem of CF, often referred to as memory-based CF, it is to compute similarity between users or items on the entire collection of previously rated items by the users \cite{resnick1994grouplens}. Depending on how the neighborhood is defined, memory based CF comes in two flavors: user-based or item-based CF. Although, the term ``collaborative filtering'' is commonly used as a synonym for user-based CF \cite{Ziegler2013}, the two approaches share the same common abstract formulation: both deal with the prediction of unassigned elements of a given matrix relying on the statistics of elements already assigned. Rows and columns take the role of users and items, while matrix's elements represent rates given by the users on items. Within this abstract view, memory-based CF exploits similarities across a population of vectors from the matrix above by forming weighted rates to predict unrated items.

This observation provides the basis for a link between CF and information retrieval (IR). 
For instance, user-based CF relies on the calculation of the similarity between users almost in the same way as IR systems calculate the similarity between documents. Both their algorithms essentially calculate the dot product between vectors, which take the form of user preferences, in the case of CF, and documents, in the case of IR.

The work by Breese et al. \cite{breese1998empirical} is definitely the first research that has found connections between CF and IR. It used the cosine similarity in a vector space model, in which users take the role of documents,
titles take the role of words, and rates take the role of word frequencies. Although, this first attempt is only valid for the user-based CF approach, it can be easily extended to the case of item-based CF by replacing the users
with the items. In \cite{Coster:2002:IFS:564376.564420}, the authors propose the use of an inverted index (the core of any IR system) for user-based CF in order to speed up the execution time. However, they had to build a CF system from scratch, which is in general challenging and resource demanding.


This fact suggests that the idea of building a CF on top of a conventional IR system is not so absurd. Unfortunately, in spite of the similarity between the two approaches, there are essential differences in the way in which the weights of the vectors are eventually calculated. In fact, IR systems typically use the well--known tf-idf weighing scheme. Moreover, in memory based CF, one difficulty in measuring the user-user similarity is that the raw ratings may contain biases caused by the different behaviors of different users. For example, some users
may tend to give high ratings \cite{liu2008eigenrank}. To correct such biases, one solution is to ``mean-center'' the user rates before computing the cosine similarity, which corresponds to consider the Pearson correlation.

To this end, we propose an approach to CF, which uses a document-based representation of users/items and their rates so that when documents are indexed by a conventional retrieval system, such as Lucene (\texttt{http://lucene.apache.org}), the vector weights assume the form of mean-centered rates.

A similar approach to the one we propose can be found in \cite{bellogin2013bridging}.
However, this approach to CF has the disadvantage of using a cumbersome transformation technique in which for each item $i$ (represented as document) they store in an inverted index the items (now acting as terms) most
similar to $i$. Once the inverted index has been built, a conventional text retrieval engine needs two more operations (called query transformation and processing) in order to generate a document ranking for a particular query.

As it will be shown, our approach to CF is much more practicable and flexible. We exploit Lucene features to build an index from the free movie recommender site MovieLens in which documents representing user profiles are stored.
An experimental evaluation shows that our method produces results that are even better than those that would obtain using the state--of--art techniques for CF.


\section{Memory-Based CF}

Let $U$ be a set of $N$ users, $I$ a set of $M$ items, and let $r_{u,i}$ denote the rating of users
$u \in U$ on item $i \in I$. $S_u \subseteq I$ stands for the set of items that user $u$ has rated and $S_i \subseteq U$ is the set of users that have rated item $i$. The purpose of CF is to predict the rating $p_{u,i}$ of a given user $u$ on a given item $i$. User $u$ is supposed to be active,
meaning that he/she has already rated some items, therefore $S_u \neq \emptyset$. But the item to be
predicted shall not be known by the user, so $i \notin S_u$ \cite{Candillier2008}.

In user-based CF, we compute similarity between users. To accomplish this task the $k$-Nearest-Neighbors (kNN) of a target user's profile with the historical profiles of other users are retrieved. The recommendation is then made by aggregating the ratings of these neighbors. In item-based CF, we similarly look for items sharing common user preferences. Therefore, in this case, we compute the kNN of a target item.

\subsection{Similarity Functions}\label{subsec:sim}

Several similarity functions have been proposed for measuring both the similarity between users or items to find the kNN. These functions are sometimes referred to as \emph{correlation coefficients}. Probably the most popular and also the oldest one is the Pearson correlation introduced in \cite{resnick1994grouplens}. It corresponds to the cosine of rates' deviations from the mean, which for users-based approach is given by:
{\scriptsize
$$ sim_P(u,a) = \frac{ \sum_{i\in S_u \bigcap S_a}(r_{u,i} - \overline{r}_{u,a})(r_{a,i} - \overline{r}_{a,u})}{\sqrt{ \sum_{i\in S_u \bigcap S_a}(r_{u,i} - \overline{r}_{u,a})^2} \sqrt{ \sum_{i\in S_u \bigcap S_a}(r_{a,i} - \overline{r}_{a,u})^2}}$$
}
where $
\overline{r}_{u,a} = (\sum_{i\in S_u \bigcap S_a} r_{u,i})/|S_u \bigcap S_a|
$. While, in the item-based approach, we have:\par
{\scriptsize
$$
sim_P(i,j) = \frac{\sum_{a\in S_i \bigcap S_j}(r_{a,i} - \overline{r}_{i,j})(r_{a,j} - \overline{r}_{j,i})}{\sqrt{\sum_{a\in S_i \bigcap S_j}(r_{a,i} - \overline{r}_{i,j})^2} \sqrt{\sum_{a\in S_i \bigcap S_j}(r_{a,j} - \overline{r}_{j,i})^2}}
$$
}
where
$
\overline{r}_{i,j} = (\sum_{a\in S_i \bigcap S_j} r_{a,i})/|S_i\bigcap S_j|
$.
Another popular similarity function is the adjusted cosine similarity (ACS), $sim_A$, introduced by \cite{sarwar2001item}. It corresponds to the cosine of the angle between the user/item rating vectors, after adjusting each rating by subtracting their average rating (mean-centered rates). For user-based approach $sim_A$ is defined as:\par
{\scriptsize
\begin{equation}\label{eq:uacs}
sim_A(u,a) = \frac{\sum_{i\in S_u \bigcap S_a}(r_{u,i} - \overline{r}_u)(r_{a,i} - \overline{r}_a)}{\sqrt{\sum_{i\in S_u \bigcap S_a}(r_{u,i} - \overline{r}_u)^2} \sqrt{\sum_{i\in S_u \bigcap S_a}(r_{a,i} - \overline{r}_a)^2}}
\end{equation}
} where
$
\overline{r}_{u} = (\sum_{i\in S_u} r_{u,i})/|S_u|
$.
While, in the item-based approach, we have:\par
{\scriptsize
\begin{equation}\label{eq:iacs}
sim_A(i,j) = \frac{\sum_{a\in S_i \bigcap S_j}(r_{a,i} - \overline{r}_i)(r_{a,j} - \overline{r}_j)}{\sqrt{\sum_{a\in S_i \bigcap S_j}(r_{a,i} - \overline{r}_i)^2} \sqrt{\sum_{a\in S_i \bigcap S_j}(r_{a,j} - \overline{r}_j)^2}}
\end{equation}
}
where\par
$
\overline{r}_{i} = (\sum_{a\in S_i} r_{a,i})/|S_i|.
$
\par

%

\subsection{Predicting Unrated Items}

Now that we have defined the notion of similarity between users and items, we show how to predict the rating of an unrated item.
In memory-based approach to CF, 
the similarity function can be seen as ``plug-in'' component of the filtering algorithm as discussed above. For predicting the rating of user $a$ on item $j$, user-based CF uses the average rating $\overline{r}_a$, and the set of kNN ($\mathcal{N}_a$) to user $a$ as follows:
{\scriptsize
\begin{equation}\label{eq:predict-user}
    p_{a,j} =  \overline{r_a} + \frac{\sum_{u \in \mathcal{N}_a} sim(a,u) \times (r_{u,j}-\overline{r}_{u})}{\sum_{u \in \mathcal{N}_a} |sim(a,u)| }
\end{equation}
}
In general this type of estimate can be thought of as the sum of two contributions called \emph{base estimate} and \emph{neighborhood estimate} \cite{Rafter2009}. Where the former contribution can be taken as a minimal estimate.

Item-based CF is the dual approach to the user-based approach and uses the similarities between items rather than similarities between users. As it is easy to see, we can use again a technique based on kNN ($\mathcal{M}_i$) of item $i$ just as we did for the user-based approach. We predict the rating of user $a$ on item $i$ as follows:
{\scriptsize
\begin{equation}\label{eq:predict-item}
    p_{a,i} =  \overline{r_i} + \frac{\sum_{j \in \mathcal{M}_i} sim(i,j) \times (r_{a,j}-\overline{r}_j)}{\sum_{j \in \mathcal{M}_i} |sim(i,j)| }
\end{equation}
}
In this paper, we focus on the ACS, since it is shown to be the most accurate similarity measure for memory-based CF systems \cite{sarwar2001item}. However, the main reason why we use ACS is related to the way in which the means are evaluated and to the idea of using a conventional text search engine as the basic tool to support CF. Let us take the user-based approach as example. Both Pearson and ACS for two users $u$ and $a$ are computed just over the ratings in $ S_u \bigcap S_a$ because we want to consider only the co-rated items of the users $u$ and $a$. However, the ``global'' average rating $\overline{r}_{u}$ and $\overline{r}_{a}$ of the adjusted cosine are computed considering all the items rated by $u$ and $a$. While, the mean $\overline{r}_{u,a}$ in Pearson correlation is the average rating of user $u$ for all the items in $S_u \bigcap S_a$. Therefore, in general $\overline{r}_{u,a} \neq \overline{r}_{a,u}$. 
The advantage of using the global average rating $\overline{r}_{u}$ is that it can be easily evaluated and stored at indexing time since it refers to a single user only. The average rate $\overline{r}_{a,u}$, used in Pearson correlation, should be evaluated at run-time, since it considers only rates of user $a$ on items that are in $S_a \bigcap S_s$. The same reasoning applies to item-based case.\par


\section{CF with Lucene} \label{sec:lucene}

Let us now enter the core of our method. In what follows, our aim will be to employ a text retrieval, such as Lucene, for easily developing a memory-based CF framework. To this end, in case of user-based CF, we can adopt a transformation in which users take the role of documents, items take the role of words, and rates take the role of word frequencies \cite{breese1998empirical}. Similarly, for item-based CF, we could let users take the role of words, items take the role of documents, and rates take the role of word frequencies. Since in standard text retrieval the similarity between two documents is often measured by treating each document as a vector of word frequencies and computing the cosine similarity by the two vectors, we can exploit the above transformation to compute similarity between users or items. This basic approach would allow us to retrieve kNN documents (i.e., users or items) on the basis of their cosine similarity ($sim_C$) as follows \par
{\scriptsize
\begin{equation}\label{eq:cs}
sim_C(a,u) = \frac{\sum_{i\in S_a \bigcap S_u}f_{a,i}f_{u,i}}{\sqrt{\sum_{i\in S_a \bigcap S_u}f_{a,i}^2} \sqrt{\sum_{i\in S_a \bigcap S_u}f_{u,i}^2},}
\end{equation}
}
where $f_{u,i}$ now represents the frequency of word $i$ in document $u$ (which in this case is equal to $r_{u,i}$).
An analogous formula can be written for the item-based case between two items $i$ and $j$. However, as proved in \cite{sarwar2001item} ACS performs best among other similarities in the CF by eliminating biases caused by the different rating behaviors of different users. Our goal therefore is to obtain the ACS on top of the text retrieval library Lucene. To this end, for user-based CF, a first approach is to make the frequency of word $i$ in the document $u$ proportional to its mean-centered rates, i.e. the term $r_{u,i} - \overline{r}_u$. For the item-based CF, the document $i$ represents an item, therefore the frequency of a word $u$ will be proportional the term $r_{u,i} - \overline{r}_i$. We call the terms $r_{u,i} - \overline{r}_u$ and $r_{u,i} - \overline{r}_i$ Mean-Centered Rates (MCRs). Word frequencies are proportional to MCRs because generally they are decimal numbers, but frequencies are integers. Therefore, we can multiply all the MCRs by a common integer factor (say 100). \par
For example, in case of user-based CF, if the rate given by the user $u$ to the item with id 37 is 4, and the average value for the user is 3.95, then the MCR will be $4-3.95 = 0.05$. We set the frequency of the word corresponding to that user to $0.05\times 100 = 5$. Note that by multiplying all the MCRs for the same factor 100, we do not affect the efficacy of the ACS because the frequencies are eventually normalized (apart from the truncation to two decimal places). Finally, the document corresponding to user $u$ will contain the movie id 37 repeated five times, i.e., ``37 37 37 37 37''. It is easy to see that by substituting the words frequencies (which are proportional to the MCRs) in place if the user rates in Eq. (\ref{eq:cs}), we obtain the formula of the ACS. \par
However, we still have to face another problem: how to deal with negative values of MCRs. To this end, we group the MCRs of a document into two categories: positive and negative MCRs. We refer the former group and the latter group as pMCRs and nMCRs, respectively. Positive contributions to the numerator of the ACS in Eq. (\ref{eq:uacs}) are given by the product of pMCRs with themselves plus the product of the nMCRs with themselves. Therefore, if we associate separate ids to the pMCRs and nMCRs, these will contribute to the cosine similarity independently, exactly as described in Eq. (\ref{eq:uacs}). Continuing with our example of user-based CF, suppose we have another item of id, (say 24) corresponding to a nMCRs = -0.04. We decide then to use the prefix 'p' for the id corresponding to the pMCRs while we use the prefix 'n' for the nMCRs. The content of the document (i.e., the user $u$) therefore will be ``p37 p37 p37 p37 p37 n24 n24 n24 n24''. Suppose we have the following query, ``p37 p37 p37 p24 p24 p24 p24 p24 p24'', which corresponds to a query user $a$ of which we want to calculate the similarity with the above user. The matching function (standard cosine similarity) of the text retrieval system will compute the following similarity due to the contribution of p37 in both users:\par
{\scriptsize
$$sim_C(a,u) = \frac{f_{a,37}*f_{u,37}}{\sqrt{f_{a,37}^2+f_{a,24}^2}\sqrt{f_{u,37}^2+f_{u,24}^2}} = \frac{3*5}{\sqrt{3^2+6^2}\sqrt{5^2+4^2}}$$
}
Using this technique, we can guarantee that the text retrieval system will automatically calculate the positive contributions of the ACS during the simple cosine similarity calculation. But what about the negative contributions due to the cross-products between pMCRs and nMCRs? Lucene's fields can aid us in solving this problem. Document corresponding to a user can be structured in two separate fields, one to calculate the matches for the positive contributions (pMCR-pMCR and nMCR-nMCR), as already seen, and a second field to calculate the matches for negative contributions (pMCR-nMCR and nMCR-pMCR). The second field will just have the word with prefix 'n' replaced with the prefix 'p' and vice versa. We refer to the former fields as 'PRATE' and the latter field 'NRATE'. We then execute two queries: one on the field PRATE and the second query on the field NRATE. We merge the result sets from the two queries by subtracting the score from the query on PRATE with the score from NRATE for the same documents, from which we get the following similarity:\par
{\scriptsize
$$sim_C(a,u) = \frac{f_{a,37}*f_{u,37}-f_{a,24}*f_{u,24}}{\sqrt{f_{a,37}^2+f_{a,24}^2}\sqrt{f_{u,37}^2+f_{u,24}^2}} = \frac{3*5-6*4}{\sqrt{3^2+6^2}\sqrt{5^2+4^2}}$$
}
In order to increase the chances of getting at least $k$ results from the merged result set, the depth $k$ of the single kNN queries is increased of a constant expansion factor $m$. In item-based CF case we proceed in a similar way: the documents take the place of the items and words refer to users.

Now that we obtained that the similarity between documents is identical to that which we obtained using Eq. (\ref{eq:uacs}) and (\ref{eq:iacs}), we can use the scores obtained by the ACS to predict the missing rate using Eq. (\ref{eq:predict-user}) and (\ref{eq:predict-item}), respectively.

\section{Experimental Evaluation}
\label{sec:experimental}
Our experiments have been carried out using the public dataset Movielens \cite{resnick1994grouplens}. 
Among the three available datasets of Movielens 100K, 1M, and 10M, we decided the use the biggest one, which contains 10,000,054 ratings applied to 10,681 movies by 71,567 users.
Starting from the initial data set, five distinct splits of training and test data were generated, the training set (80\%) and the test set (20\%). The test set acts as a ground-truth for our validation.

To assess the statistical accuracy, we use the mean absolute error (MAE) as the measure of effectiveness, which is given by $MAE = (\sum^N_{i=1} |p_i-r_i|)/N$. 
Finally, since in our dataset the ratings are quantized, we chose to round the predicted ratings instead of using their real values. Such a process improves the performance of the prediction.

\begin{figure}[h]
    \centering
    \includegraphics[width=0.49\textwidth]{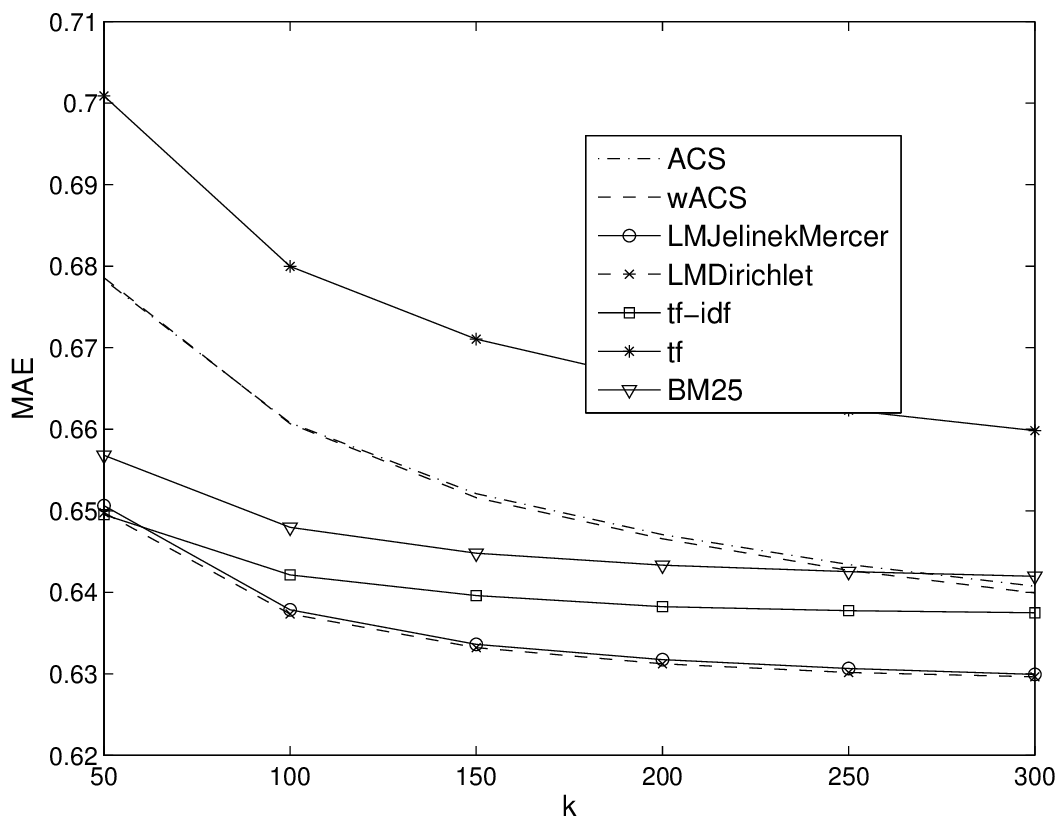}
    \includegraphics[width=0.49\textwidth]{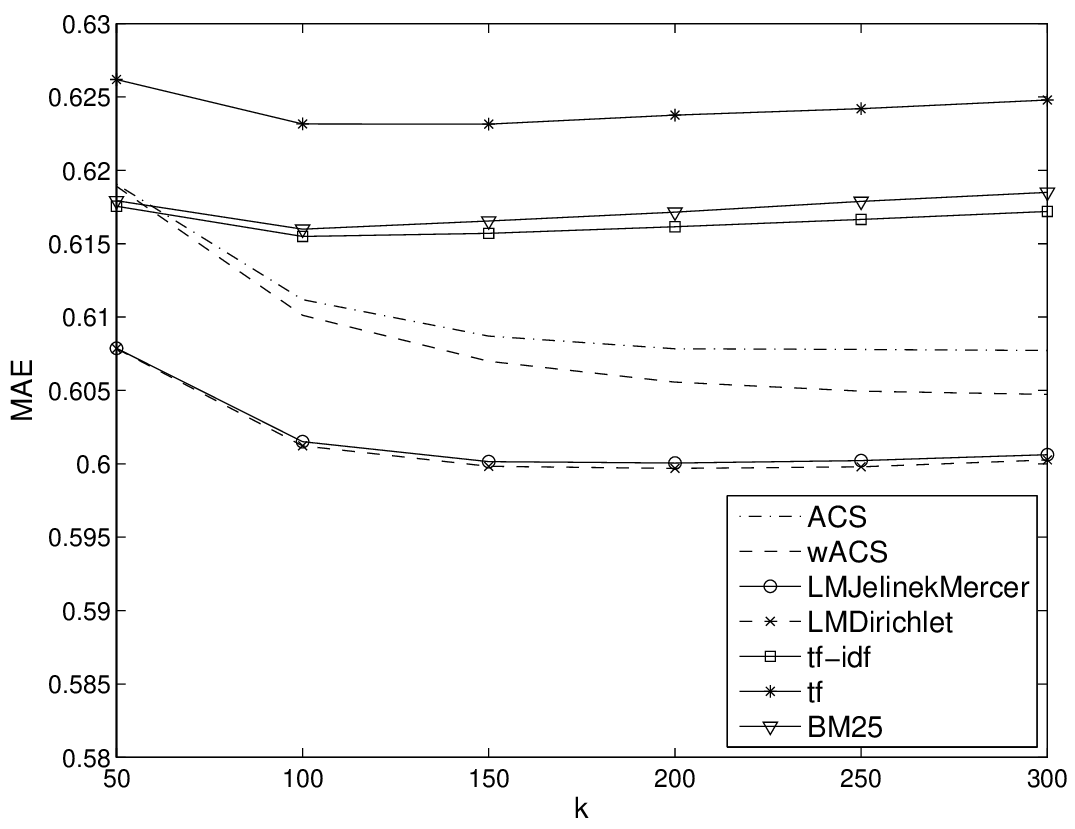}
    \caption{Comparing prediction schemes for user-based approaches (top) and item-based approaches (bottom) and different neighborhood sizes (k).}
    \label{fig:iubased}
\end{figure}

Figure \ref{fig:iubased} reports the results of user-based and item-based CF approaches for different vales of $k$. In this evaluation, we compared
our approach with the results obtained with a standard approach to CF.
In particular, we performed two additional groups of experiments in which the pairwise similarity matrix between users (for user-based approach) and between items (for item-based approach) have been computed offline. We used the best-known similarity measure, i.e., ACS.
However, since Pearson correlation weighted by Jaccard index has been shown to provide better results than the traditional Pearson coefficients alone \cite{Candillier2008}, we also boosted ACS with Jaccard coefficient. This improves the performance of ACS particularly for the item-based approach. We call this approach wACS.
The other five graphs in Figure \ref{fig:iubased} correspond to the experiment of our approach using Lucene. In particular, \emph{tf} similarity corresponds to the results obtained with the standard cosine similarity, while \emph{tf-idf} uses the product of terms frequency and inverse document frequency. We also tried the language models based on the \emph{Dirichlet} and \emph{Jelinek-Mercer} similarities \cite{zhai2001study}, and the \emph{BM25} similarity function derived using the classic probabilistic retrieval model \cite{robertson1999okapi}. In all experiments, we used $m = 10$ as query expansion factor. We repeated measures with ten different random partitions of tests/training sets in order to guarantee a confidence level greater than equal to 95\%.

According to these experiments, the similarity measures based on Dirichlet and Jelinek-Mercer lead to the best results even versus ACS and wACS correlations on the item-based approach. Moreover, in the user-based approach, MAE improves as $k$ increases. Our solution outperforms state-of-the-art methods currently available (e.g., see \cite{Ekstrand:2011:RRR:2043932.2043958}).





\section{Conclusion}
\label{sec:conclusion}
We presented a recommendation framework developed using the Lucene open--source platform. There are numerous advantages to proceeding in this manner, as opposed to developing ad-hoc solutions for CF. To name
a few: Lucene provides high performance, robust, and scalable indexing and retrieval platform that is designed to cope with Web-scale data and usage. Moreover, Lucene can be also used to store information of the users and the items over which to make recommendations. Lucene provides, fielded searching (e.g. title, author, contents).
Surprisingly, we found that with our technique the IR language models exhibited a better performance. To accelerate the prediction we use an extra field of float type to store the precomputed average user rates $\overline{r}_{u}$ or $\overline{r}_{i}$.
The two Lucene DBs take more or less the same storage space, 86.6MB for the item-based DB and 78.8MB for the item-based DB. The prediction time, including the queries, for both approaches are comparable and about 0.14sec. 

Building on Hebbian principles \cite{lagani2021hebbian}, where associations strengthen through co-activation, collaborative filtering systems could benefit from learning dynamically from user interactions. This approach would enable recommendations to adapt continuously to shifting user preferences and context, offering an intuitive and self-organizing framework for personalization.

\bibliographystyle{abbrv}
\bibliography{arxiv}

\end{document}